\documentclass[a4paper,12pt]{article}

 \usepackage{amsfonts}
\usepackage{verbatim}

\begin{document}


\vskip 0.8cm

\begin{center}

{\large \bf Notoph Gauge Theory: Superfield Formalism}

\vskip 2.5cm

{\bf R. P. Malik}\\
{\it Physics Department, Centre of Advanced Studies,}\\
{\it Banaras Hindu University, Varanasi-221 005, (U. P.), India}\\

and\\

{\it DST Centre for Interdisciplinary Mathematical Sciences,}\\
{\it Banaras Hindu University, Varanasi-221 005 (U.P.), India}\\
{\small  E-mails: rudra.prakash@hotmail.com ; malik@bhu.ac.in}\\

\end{center}

\vskip 2.5cm

\noindent
{\bf Abstract:}
We derive absolutely anticommuting Becchi-Rouet-Stora-Tyutin (BRST)
and anti-BRST symmetry transformations for the 4D free Abelian 2-form
gauge theory by exploiting the superfield approach
to BRST formalism. The antisymmetric tensor gauge field of the above theory
was christened as the ``notoph'' (i.e. the opposite of ``photon'') gauge field by Ogievetsky 
and Palubarinov way back in 1966-67. We briefly outline the  
problems involved in obtaining the absolute anticommutativity
of the (anti-) BRST transformations and their resolution
within the framework of geometrical superfield approach to BRST formalism. One of
the highlights of our results is the emergence of a Curci-Ferrari
type of restriction in the context of 4D Abelian 2-form (notoph) gauge theory which renders
the nilpotent (anti-) BRST symmetries of the theory to be absolutely 
anticommutative in nature.  \\



\newpage

\section{Introduction}

The Becchi-Rouet-Stora-Tyutin (BRST) and anti-BRST symmetry
transformations emerge when the ``classical'' local gauge symmetry transformations
of any arbitrary $p$-form ($p = 1, 2, 3......$) gauge theory are elevated to the 
``quantum'' level. The above (anti-) BRST symmetry transformations are found to be nilpotent
of order two and they anticommute with each-other. These properties are very sacrosanct
as they encode (i) the fermionic nature of these symmetries, and (ii) the linear
independence of these transformations (see, e.g. [1]). These statements are true for the BRST
approach to any arbitrary $p$-form gauge theories in any arbitrary dimension of spacetime.

In recent years, the Abelian 2-form (i.e. $B^{(2)} = (1/2!) (dx^\mu \wedge dx^\nu) B_{\mu\nu}$)
gauge theory with antisymmetric ($B_{\mu\nu} = - B_{\nu\mu}$) tensor potential\footnote{This
potential was christened as the notoph gauge field by Ogievetsky and Palubarinov 
who were the first to discuss about this gauge theory at BLTP, JINR, Dubna [2].} $B_{\mu\nu}$   
has become quite popular because of its relevance in the context of (super) string theories [3,4]
and the noncommutativity associated with them due to the
 presence of $B_{\mu\nu}$ in the background [5]. It has been shown, furthermore, that the Abelian
2-form (notoph) gauge theory provides  a tractable field theoretic model for the Hodge theory [6-8]
as well as quasi-topological field theory [9]. This theory has been discussed within the 
framework of the BRST formalism, too (see, e.g. [10-12]).
The known nilpotent (anti-) BRST
transformations, however, have been found to be anticommuting only up to the U(1)
vector gauge transformations (see, e.g. [6,7] for details).

One of the key problems in the context of 4D notoph gauge theory has been to obtain
a set of (anti-) BRST symmetry transformations that are consistent with the basic
tenets of BRST formalism. The central theme of our presentation is to obtain
{\it absolutely} anticommuting off-shell nilpotent (anti-) BRST symmetry transformations
for the notoph gauge theory by exploiting the geometrical superfield approach 
to BRST formalism proposed by Bonora and Tonin [13,14]. We demonstrate that a 
Curci-Ferrari (CF) type restriction emerges from the superfield formalism which
enables us to derive (i) the absolute anticommutativity of the (anti-) BRST symmetry 
transformations, and (ii) the coupled Lagrangian densities of the theory that respect
these (anti-) BRST symmetry transformations. The idea of the horizontality condition (HC)
is at the heart of these derivations.

The layout of our presentation  is as follows. First, we recapitulate the bare
essentials of the nilpotent (anti-) BRST symmetry transformations [6,7] that are 
anticommuting only up to a vector U(1) gauge transformation. We describe, after this, 
the key issues associated with the HC within the framework of the superfield formalism.
Next we derive the CF type restriction by exploiting the celebrated HC. The former turns
out to be (anti-) BRST invariant quantity and it leads to
the derivation of the coupled Lagrangian densities for the notoph gauge theory. These
Lagrangian densities, in turn, respect the off-shell nilpotent and absolutely anicommuting
(anti-) BRST symmetry transformations. Finally, we provide geometrical interpretations
for the nilpotent and anticommuting symmetries (and corresponding generators)
within the framework of superfield approach to BRST formalism.

\section{Preliminaries: Old Lagrangian formulation and off-shell nilpotent symmetries}

We begin with the generalized version of the Kalb-Ramond Lagrangian
density (${\cal L}^{(0)} = \frac{1}{12} H^{\mu\nu\kappa} H_{\mu\nu\kappa}$)
for the  4D\footnote{We choose, for the whole body of our present text, 
the 4D flat metric $\eta_{\mu\nu}$ with signature $(+1, -1, -1, -1)$
where the Greek indices $\mu, \nu, \eta...= 0, 1, 2, 3$. 
The convention $(\delta B_{\mu\nu}/ \delta B_{\eta\kappa})
= \frac{1}{2!} (\delta_{\mu\eta} \delta_{\nu\kappa} - \delta_{\mu\kappa} \delta_{\nu\eta})$
has been adopted in our full text [6,7].}
notoph gauge theory that respects the off-shell nilpotent
(anti-) BRST transformations [6,7]. This Lagrangian density, in its full blaze of glory, is 
as follows (see, e.g. [6,7] for details)
\begin{eqnarray}
&&{\cal L}^{(0)}_B = \frac{1}{12} H^{\mu\nu\kappa} H_{\mu\nu\kappa} 
 + B^\mu (\partial^\nu B_{\nu\mu} - \partial_\mu \phi) - \frac{1}{2} B \cdot B
- \partial_\mu \bar\beta \partial^\mu \beta \nonumber\\
&&+ (\partial_\mu \bar C_\nu - \partial_\nu \bar C_\mu) (\partial^\mu C^\nu) +
\rho \;(\partial \cdot C + \lambda) + (\partial \cdot \bar C + \rho) \;\lambda,
\end{eqnarray}
where $B_\mu = \partial^\nu B_{\nu\mu} - \partial_\mu \phi$ is the  
Lorentz vector auxiliary field that has been invoked
to linearize the gauge-fixing term, the massless ($\Box \phi = 0$) scalar field $\phi$ has
been introduced
for the stage-one reducibility in the theory and the totally antisymmetric curvature
tensor $H_{\mu\nu\kappa} = \partial_\mu B_{\nu\kappa} + \partial_\nu B_{\kappa\mu}
+ \partial_\kappa B_{\mu\nu}$ is constructed with the help of the notoph
gauge field $B_{\mu\nu}$.

The fermionic (i.e. $C_\mu^2 = \bar C_\mu^2 = 0, C_\mu \bar C_\nu + \bar C_\nu C_\mu = 0 $, etc.,)
Lorentz vector (anti-) ghost fields
$(\bar C_\mu)C_\mu$ (carrying ghost number $(-1)1$) have been introduced 
to compensate for the above gauge-fixing term and they
play important roles in the existence of the 
(anti-) BRST symmetry transformations for the notoph gauge potential.
The bosonic (anti-) ghost fields
$(\bar\beta)\beta$ (carrying ghost numbers $(-2)2$) are needed for
the requirement of 
ghost-for-ghost in the theory. The auxiliary ghost fields 
$\rho = - \frac{1}{2} (\partial \cdot \bar C)$ and $\lambda = - \frac{1}{2}
(\partial \cdot  C)$ (with ghost numbers (-1)1) are also present in the theory.

The following off-shell nilpotent ($\tilde s_{(a)b}^2 = 0$) (anti-) BRST
symmetry transformations $\tilde s_{(a)b}$ for the fields of the Lagrangian
density (1):
\begin{eqnarray}
&& \tilde s_b B_{\mu\nu} = - (\partial_\mu C_\nu - \partial_\nu C_\mu), \qquad 
\tilde s_b C_\mu  = - \partial_\mu \beta, \qquad \tilde s_b \bar C_\mu = - B_\mu,
\nonumber\\
&& \tilde s_b \phi = \lambda, \qquad \tilde s_b \bar \beta = - \rho, \qquad
\tilde s_b [\rho, \lambda, \beta, B_\mu,  H_{\mu\nu\kappa}] = 0, 
\nonumber\\
&& \tilde s_{ab} B_{\mu\nu} = - (\partial_\mu \bar C_\nu - \partial_\nu \bar C_\mu), \quad 
\tilde s_{ab} \bar C_\mu  = + \partial_\mu \bar \beta, \quad \tilde s_{ab}  C_\mu = + B_\mu,
\nonumber\\
&& \tilde s_{ab} \phi = \rho, \qquad \tilde s_{ab}  \beta = - \lambda, \qquad
\tilde s_{ab} [\rho, \lambda, \bar \beta, B_\mu,  H_{\mu\nu\kappa}] = 0, 
\end{eqnarray}
leave the Lagrangian density (1) quasi-invariant because it changes to the total spacetime
derivatives as given below:
\begin{eqnarray}
\tilde s_b {\cal L}^{(0)}_B &=& - \partial_\mu \bigl [ B^\mu \lambda + 
(\partial^\mu C^\nu - \partial^\nu C^\mu) B_\mu - \rho \partial^\mu \beta \bigr ], \nonumber\\
\tilde s_{ab} {\cal L}^{(0)}_B &=& - \partial_\mu \bigl [ B^\mu \rho + 
(\partial^\mu \bar C^\nu - \partial^\nu \bar C^\mu) B_\mu - \lambda \partial^\mu \bar \beta \bigr ].
\end{eqnarray}
Thus, the action corresponding to the Lagrangian density (1) remains invariant under 
the off-shell  nilpotent (anti-) BRST transformations (2).

It can be checked that $(\tilde s_b \tilde s_{ab} + \tilde s_{ab} \tilde s_b) C_\mu 
= \partial_\mu \lambda$ and $(\tilde s_b \tilde s_{ab} + \tilde s_{ab} \tilde s_b) \bar C_\mu 
= - \partial_\mu \rho$.  The above anticommutator
for the rest of the fields, however, turns out to be absolutely zero.
Thus, we note that the (anti-)BRST transformations are anticommuting only up to the U(1) vector
gauge transformations. They are {\it not} absolutely anticommuting for fields $C_\mu$ and 
$\bar C_\mu$.  In other words, the off-shell nilpotent 
(anti-) BRST symmetry transformations (2) are {\it not}
consistent with the basic tenets of BRST formalism.

\section{Horizontality condition: A synopsis}

The off-shell nilpotency and absolute anticommutativity properties of the (anti-) BRST symmetry
transformations are the natural consequences of the application of the superfield approach
to BRST formalism [13-16]. Thus, we take recourse to this
formalism to resolve the problem that has been stated earlier. In fact, we
derive the off-shell nilpotent and absolutely anticommuting (anti-) BRST symmetry transformations
for the notoph gauge theory by exploiting the celebrated horizontality condition
within the framework of the geometrical superfield approach [13-16].

The notoph gauge theory is endowed with the first-class constraints [17] in the language
of Dirac's prescription for the classification scheme. As a consequence, the theory
respects a local gauge symmetry transformation that is generated by these constraints.
The above classical local symmetry transformation is traded with the (anti-) BRST symmetry transformations
at the quantum level. The latter can be derived by exploiting the superfield formalism [18].
One of the key ingredients in the superfield formulation is to consider the 4D ordinary gauge theory
on a (4, 2) dimensional supermanifold where one has the following $N = 2$ generalizations [18]
\begin{eqnarray}
&& x^\mu \to Z^M = (x^\mu, \theta, \bar \theta), \quad d = dx^\mu \partial_\mu \to
\tilde d = dx^\mu \partial_\mu + d \theta \partial_\theta + d \bar\theta \partial_{\bar\theta}, \nonumber\\
&& B^{(2)}  = \frac{1}{2!} (dx^\mu \wedge dx^\nu) B_{\mu\nu} 
\to \tilde B^{(2)} = \frac{1}{2!} (dZ^M \wedge dZ^N) B_{MN}. 
\end{eqnarray}
In the above, $Z^M = (x^\mu, \theta, \bar\theta)$ is the $N = 2$ superspace variable, $\theta$ and $\bar\theta$
are the Grassmannian variables (with $\theta^2 = \bar\theta^2 = 0, \theta \bar\theta + \bar\theta \theta = 0$),
$\tilde d = d Z^M \partial_M$ is the super exterior derivative (with $\partial_M = (\partial_\mu, \partial_\theta,
\partial_{\bar\theta})$) and $\tilde B^{(2)}$ is the super 2-form (notoph) gauge field 
connection with a few multiplet superfields.

The explicit form of the above super 2-form connection field is as follows 
\begin{eqnarray}
\tilde B^{(2)} &=& \frac{1}{2!} (dx^\mu \wedge dx^\nu) \; \tilde B_{\mu\nu} (x, \theta, \bar\theta)
+ (dx^\mu \wedge d \theta)\; \tilde {\bar {\cal F}_\mu} (x, \theta, \bar\theta) \nonumber\\
&+& (dx^\mu \wedge d \bar\theta) \;\tilde {\cal F}_\mu (x, \theta, \bar\theta)
+ (d \theta \wedge d \theta) \; \tilde {\bar \beta} (x, \theta, \bar\theta) \nonumber\\
&+& (d \bar\theta \wedge d \bar \theta) \; \tilde \beta (x, \theta, \bar\theta)
+ (d\theta \wedge d \bar\theta) \;\tilde \Phi (x, \theta, \bar\theta).
\end{eqnarray}
In the above, the (4, 2)-dimensional multiplet superfields (see, e.g. [18]) 
\begin{eqnarray}
\tilde B_{\mu\nu} (x, \theta, \bar\theta), \;\tilde {\bar {\cal F}_\mu} (x, \theta, \bar\theta),\;
\tilde {\cal F}_\mu (x, \theta, \bar\theta), \;\tilde {\bar \beta} (x, \theta, \bar\theta),\;
\tilde \beta (x, \theta, \bar\theta), \;\tilde \Phi (x, \theta, \bar\theta),
\end{eqnarray}
are the generalizations of the basic local fields $B_{\mu\nu}, \bar C_\mu, C_\mu, \bar \beta, \beta, \phi$
of the nilpotent (anti-) BRST invariant Lagrangian density (1) of the 4D notoph gauge theory. This can be
explicitly seen by the following super expansion of these superfields along the Grassmannian
directions of the supermanifold:
 \begin{eqnarray}
\tilde {\cal B}_{\mu\nu} (x, \theta, \bar\theta) &=& B_{\mu\nu}
(x) + \theta\; \bar R_{\mu\nu} (x) + \bar\theta\; R_{\mu\nu} (x) +
i \;\theta \; \bar\theta\; S_{\mu\nu} (x), \nonumber\\ \tilde \beta
(x, \theta, \bar\theta ) &=& \beta (x) + \theta \;\bar f_1 (x) +
\bar\theta\; f_1 (x) + i\; \theta\; \bar\theta\; b_1 (x),
\nonumber\\ \tilde {\bar \beta} (x, \theta, \bar\theta) &=& 
\bar\beta (x) + \theta \;\bar f_2 (x) + \bar \theta\; f_2 (x) + i
\;\theta\;\bar\theta\; b_2 (x), \nonumber\\ \tilde \Phi (x,
\theta, \bar\theta) &=& \phi (x) + \theta \;\bar f_3 (x) +
\bar\theta\; f_3 (x) + i \;\theta \;\bar\theta\; b_3 (x),
\nonumber\\ \tilde {\cal F}_\mu (x, \theta, \bar\theta) &=& C_\mu
(x) + \theta \;\bar B^{(1)}_\mu (x) + \bar\theta\; B^{(1)}_\mu (x)
+ i \;\theta \; \bar\theta\;f^{(1)}_\mu (x), \nonumber\\ \tilde
{\bar {\cal F}}_\mu (x, \theta, \bar\theta) &=& \bar C_\mu (x) +
\theta\; \bar B^{(2)}_\mu (x) + \bar\theta\; B^{(2)}_\mu (x) + i
\; \theta\; \bar\theta \bar f^{(2)}_\mu (x).
\end{eqnarray}
In the limit $(\theta, \bar\theta) \to 0$, we retrieve our basic local fields of the 
original 4D notoph gauge theory.
Furthermore, the above expansion is in terms of the basic fields (6) and 
rest of the fields in the expansion are secondary fields.

To obtain the explicit form of the secondary fields in terms of the basic fields, one has to
invoke the celebrated HC (i.e. $ \tilde d \tilde B^{(2)} = d B^{(2)}$) which is the requirement that the 
curvature 3-form $H^{(3)} = d B^{(2)}$ remains unaffected by the presence of the supersymmetry in the 
theory. In other words, all the Grassmannian components of the following super 3-form
\begin{eqnarray}
\tilde d \tilde B^{(2)} &=& {\displaystyle \frac{1}{2!}}
(dx^\kappa \wedge d x^\mu \wedge dx^\nu) (\partial_\kappa \tilde
{\cal B}_{\mu\nu}) + (d \theta \wedge d \theta \wedge d \theta)
(\partial_{\theta} \tilde {\bar \beta}) \nonumber\\ 
&+& (d\theta \wedge d\bar\theta \wedge
d\bar\theta) \;\bigl [\partial_{\bar\theta} \tilde \Phi +
\partial_\theta \tilde \beta \bigr ] +
(d\bar \theta \wedge d\theta \wedge d\theta)\; \bigl
[\partial_{\theta} \tilde \Phi +
\partial_{\bar\theta} \tilde {\bar\beta} \bigr ]\nonumber\\
&+& {\displaystyle \frac{1}{2!}} (dx^\mu \wedge dx^\nu \wedge d
\theta)\; \bigl [
\partial_{\theta} \tilde {\cal B}_{\mu\nu} + \partial_\mu
\tilde {\bar {\cal F}}_\nu -
\partial_\nu \tilde {\bar {\cal F}}_\mu \bigr ] \nonumber\\
&+& (dx^\mu \wedge d \theta \wedge d \theta) \;\bigl
[\partial_{\theta} \tilde {\bar {\cal F}}_\mu +
\partial_\mu \tilde {\bar \beta} \bigr ]
+ (dx^\mu \wedge d \bar\theta \wedge d \bar \theta) \;\bigl
[\partial_{\bar\theta} \tilde {\cal F}_\mu +
\partial_\mu \tilde  \beta \bigr ]\nonumber\\
&+& {\displaystyle \frac{1}{2!}} (dx^\mu \wedge dx^\nu \wedge d
\bar \theta)\; \bigl [
\partial_{\bar\theta} \tilde {\cal B}_{\mu\nu} + \partial_\mu
\tilde {\cal F}_\nu -
\partial_\nu \tilde  {\cal F}_\mu \bigr ] \nonumber\\
&+& (dx^\mu \wedge d\theta \wedge d\bar\theta)\; \bigl [
\partial_\mu \tilde \Phi + \partial_\theta \tilde {\cal F}_\mu +
\partial_{\bar\theta} \tilde {\bar {\cal F}}_\mu \bigr ] +
(d \bar\theta \wedge d
\bar \theta \wedge d \bar\theta) (\partial_{\bar\theta} \tilde \beta),
\end{eqnarray}
are to be set equal to zero. This condition has been referred to as the
soul-flatness condition by Nakanashi and Ojima [19].

Physically, the soul-flatness condition (or HC) is the requirement that the gauge (i.e. (anti-) BRST)
invariant quantity (i.e. curvature tensor) should remain independent of the Grassmannian coordinates that 
are present in the superspace variable $Z^M = (x^\mu, \theta, \bar\theta)$. This is evident
from equation (2) where we note that $\tilde s_{(a)b} H_{\mu\nu\kappa} = 0$.
The celebrated HC, we emphasize once again, always leads to the symmetry transformations that are nilpotent
and absolutely anticommuting because these are the properties that are associated with
 the Grassmannian variables that play very important role in HC.
We shall be able to see these consequences in the next section.

\section{Curci-Ferrari type restriction and superfield expansions: Superfield formalism}

As a consequence of the HC, we can set the coefficients of the 3-form
differentials  $(d\theta \wedge d\theta \wedge d\theta)$, 
$(d\bar\theta \wedge d\bar\theta \wedge d\bar\theta)$,
$(d\theta \wedge d\theta \wedge d\bar\theta)$,
$(d\theta \wedge d\bar\theta \wedge d\bar\theta)$ equal to zero. These
requirements lead to the following conditions on some
of the secondary fields that are present in the expansions of the superfields:
\begin{eqnarray}
f_1 = \bar f_2 = b_1 = b_2 = b_3 = 0, \qquad f_2 + \bar f_3 = 0, \qquad 
\bar f_1 + f_3 = 0. 
\end{eqnarray}
In an exactly similar fashion, setting the coefficients of the differentials
$(dx^\mu \wedge dx^\nu \wedge d\theta), (dx^\mu \wedge dx^\nu \wedge d\bar\theta),
(dx^\mu \wedge d\theta \wedge d\theta), (dx^\mu \wedge d\bar\theta \wedge d\bar\theta)$
equal to zero, we obtain the following conditions on some of the secondary fields [18]:
\begin{eqnarray}
&& B_\mu^{(1)} = - \partial_\mu \beta, \quad \bar B_\mu^{(2)} = - \partial_\mu \bar \beta,
\quad f_\mu^{(1)} = i \partial_\mu \lambda, \quad 
\bar f_\mu^{(2)} = - i \partial_\mu \rho, \nonumber\\
&& R_{\mu\nu} = - (\partial_\mu C_\nu - \partial_\nu C_\mu), \qquad
\bar R_{\mu\nu} = - (\partial_\mu \bar C_\nu - \partial_\nu \bar C_\mu), \nonumber\\
&& S_{\mu\nu} = - i (\partial_\mu B_\nu - \partial_\nu B_\mu) \equiv
- i (\partial_\mu \bar B_\nu - \partial_\nu \bar B_\mu),
\end{eqnarray}
where we have identified $\bar B_\mu^{(1)} = \bar B_\mu, B_\mu^{(2)} = - B_\mu$.

Finally, it is very interesting to point out that we obtain the 
(anti-) BRST invariant Curci-Ferrari (CF) type restriction
\begin{eqnarray}
B_\mu - \bar B_\mu - \partial_\mu \phi = 0,
\end{eqnarray}
when we set the coefficient of the 3-form differential $(dx^\mu \wedge d \theta \wedge d \bar\theta)$
equal to zero (due to the celebrated HC). It would be worthwhile to state that one encounters such
kind of restriction in the case of 4D non-Abelain 1-form gauge theory [20] which enables one to obtain
the absolute anticommutativity of the off-shell nilpotent (anti-) BRST symmetry transformations. The derivation
of the CF restriction [20] within the framework of 
superfield formalism (in the context of the 4D non-Abelian 1-form gauge theory) has been performed
by Bonora and Tonin (see, e.g. [13] for details).

The stage is now set for the comparison of the coefficient of the 3-form differential
$(dx^\mu \wedge dx^\nu \wedge dx^\kappa)$ from the l.h.s. and r.h.s of the horizontality condition
$\tilde d \tilde B^{(2)} = d B^{(2)}$ where the r.h.s.
produces $\frac{1}{3!} (dx^\mu \wedge dx^\nu \wedge dx^\kappa) H_{\mu\nu\kappa}$
only. However, there are terms with Grassmannian variables on the l.h.s. Setting these
terms equal to zero leads to 
\begin{eqnarray}
\partial_\mu R_{\nu\kappa} + \partial_\nu R_{\kappa\mu} +
\partial_\kappa R_{\mu\nu} &=& 0, \qquad
\partial_\mu \bar R_{\nu\kappa} + \partial_\nu \bar R_{\kappa\mu}
+ \partial_\kappa \bar R_{\mu\nu} = 0, \nonumber\\
\partial_\mu S_{\nu\kappa} + \partial_\nu S_{\kappa\mu} + \partial_\kappa
S_{\mu\nu} &=& 0,
\end{eqnarray}
which are automatically satisfied due to values in equation (10).

Let us focus on the expansion of the superfield $\tilde {\cal B}_{\mu\nu} (x, \theta, \bar\theta)$
with the values that are given in (10). We obtain the following
\begin{eqnarray}
\tilde {\cal B}_{\mu\nu} (x, \theta, \bar\theta) &=& B_{\mu\nu} (x) 
- \theta \;(\partial_\mu \bar C_\nu - \partial_\nu \bar C_\mu) - \bar \theta \;
(\partial_\mu  C_\nu - \partial_\nu  C_\mu)\nonumber\\
&+& \theta \bar\theta \; (\partial_\mu B_\nu - \partial_\nu B_\mu).
\end{eqnarray}
Having our knowledge of the local gauge symmetry and corresponding nilpotent
(anti-) BRST symmetries, we know that the coefficient of $\theta$ in the above
is nothing but the anti-BRST symmetry transformation and that
of $\bar\theta$ is the BRST symmetry transformation. We can now guess that
the coefficient of $\theta\bar\theta$ should be the anticommutator
of (anti-) BRST symmetry transformations because of the anticommuting
properties associated with the Grassmannian 
variables. Finally, it can be seen that we have the following expansion
\begin{eqnarray}
\tilde {\cal B}^{(h)}_{\mu\nu} (x, \theta, \bar\theta) &=& B_{\mu\nu} (x) 
\;+ \;\theta \;(s_{ab} B_{\mu\nu} (x)) \;+ \;\bar \theta \;
(s_b B_{\mu\nu} (x)) \nonumber\\
&+&\; \theta \bar\theta \;(s_b s_{ab} B_{\mu\nu} (x)).
\end{eqnarray}
where the superscript $(h)$ denotes the expansion of the gauge  superfield
$\tilde {\cal B}_{\mu\nu} (x, \theta, \bar\theta)$ after the application of HC 
and symbols $s_{(a)b}$ correspond to the {\it correct} (anti-) BRST symmetry transformations
that are always nilpotent of order two and absolutely anticommuting in nature.

The substitution of all the values of the secondary fields from (10),
leads to the following expansion of the rest of the superfields  of (7), namely;
\begin{eqnarray}
\tilde \beta^{(h)} (x, \theta,
\bar\theta ) &=& \beta (x) + \theta \;(s_{ab} \beta (x)) +
\bar\theta\; (s_b \beta (x)) + \theta\; \bar\theta\; (s_b s_{ab}
\beta (x)), \nonumber\\ \tilde {\bar \beta}^{(h)} (x, \theta,
\bar\theta) &=&  \bar\beta (x) + \theta \; (s_{ab} \bar \beta
(x)) + \bar \theta\; (s_b \bar \beta (x)) + \theta\; \bar\theta\;
(s_b s_{ab} \bar \beta (x)), \nonumber\\ \tilde \Phi^{(h)} (x,
\theta, \bar\theta) &=& \phi (x) + \theta \;(s_{ab} \phi (x))
+ \bar\theta\; (s_b \phi (x)) + \theta\; \bar\theta \; (s_b
s_{ab} \phi (x)),  \nonumber\\ \tilde {\cal F}^{(h)}_\mu (x,
\theta, \bar\theta) &=& C_\mu (x) + \theta \;(s_{ab} C_\mu (x)) +
\bar\theta\; (s_b C_\mu (x)) + \theta\; \bar\theta\; (s_b s_{ab}
C_\mu (x)), \nonumber\\ \tilde {\bar {\cal F}}^{(h)}_\mu (x,
\theta, \bar\theta) &=& \bar C_\mu (x) + \theta (s_{ab} \bar
C_\mu (x)) + \bar\theta (s_b \bar C_\mu (x)) + \theta
\bar\theta  (s_b s_{ab} \bar C_\mu(x)).
\end{eqnarray}
Thus, we have obtained the absolutely anticommuting (anti-) BRST symmetry transformations
for the notoph gauge theory as 
\begin{eqnarray}
&& s_b B_{\mu\nu} = - (\partial_\mu C_\nu - \partial_\nu C_\mu), \quad 
s_b C_\mu  = - \partial_\mu \beta, \quad  s_b \bar C_\mu = - B_\mu,
\nonumber\\
&& s_b \phi = \lambda, \qquad  s_b \bar \beta = - \rho, \qquad
\tilde s_b [\rho, \lambda, \beta, B_\mu,  H_{\mu\nu\kappa}] = 0, 
\nonumber\\
&& s_{ab} B_{\mu\nu} = - (\partial_\mu \bar C_\nu - \partial_\nu \bar C_\mu), \quad 
s_{ab} \bar C_\mu  = - \partial_\mu \bar \beta, \quad  s_{ab}  C_\mu = + \bar B_\mu,
\nonumber\\
&& s_{ab} \phi = \rho, \qquad  s_{ab}  \beta = - \lambda, \qquad
\tilde s_{ab} [\rho, \lambda, \bar \beta, B_\mu,  H_{\mu\nu\kappa}] = 0, 
\end{eqnarray}
which are different from earlier nilpotent transformations (2).

It can be checked that $(s_b s_{ab} + s_{ab} s_b) B_{\mu\nu} (x) = 0$
is true if and only if we impose the CF type restriction (11) that has
emerged out from the application of superfield formalism to the notoph
gauge theory. Furthermore, the absolute anticommutativity criterion dictates
the (anti-) BRST symmetry transformations on the 
auxiliary fields $B_\mu$ and $\bar B_\mu$ as 
\begin{eqnarray}
s_b \bar B_\mu = - \partial_\mu \lambda, \quad s_{ab} B_\mu = - \partial_\mu \rho,
\quad s_b B_\mu = 0, \quad s_{ab} \bar B_\mu = 0.
\end{eqnarray}
Under the off-shell nilpotent (anti-) BRST symmetry transformations (16)
and (17), it can be seen that the absolute anticommutativity is satisfied
for all the fields of the theory which can be generically expressed as:
\begin{eqnarray}
\{s_b, s_{ab} \} \Omega = 0, \quad \Omega = C_\mu, \bar C_\mu, \beta, \bar \beta,
B_\mu, \bar B_\mu, \rho, \lambda, \phi, 
\end{eqnarray}
where $\Omega$ is the generic local field of the 4D notoph theory. 

\section{Coupled Lagrangian densities: Derivation
from (anti-) BRST approach}

With the (anti-) BRST symmetry transformations (listed in (16) and (17)),
it can be seen that the Lagrangian density for the theory can be written
in two different ways. These are as follows
\begin{eqnarray}
{\cal L}_B &=& \frac{1}{12} H^{\mu\nu\kappa} H_{\mu\nu\kappa}
+ s_b s_{ab} \Bigl [2 \beta \bar \beta + \bar C_\mu C^\mu - \frac{1}{4} B^{\mu\nu} B_{\mu\nu} \Bigr ],
\nonumber\\
{\cal L}_{\bar B} &=& \frac{1}{12} H^{\mu\nu\kappa} H_{\mu\nu\kappa}
- s_{ab} s_b \Bigl [2 \beta \bar \beta + \bar C_\mu C^\mu - \frac{1}{4} B^{\mu\nu} B_{\mu\nu} \Bigr ],
\end{eqnarray}
where the first term is nothing but the kinetic term for the notoph gauge field which is automatically
gauge (and, therefore, (anti-) BRST)) invariant. The explicit form of the bracketed terms are
\begin{eqnarray}
&& s_b\; s_{ab}\; \Bigl [ \;2 \beta \bar\beta + \bar C_\mu C^\mu - 
{\displaystyle \frac{1}{4}}
B^{\mu\nu} B_{\mu\nu} \;\Bigr ] = B^\mu (\partial^\nu B_{\nu\mu})
+ B \cdot \bar B  + \nonumber\\ && \partial_\mu \bar \beta  \partial^\mu \beta
+ (\partial_\mu \bar C_\nu - \partial_\nu \bar C_\mu) (\partial^\mu C^\nu)
+ (\partial \cdot C - \lambda)  \rho + (\partial \cdot \bar C + \rho) \lambda,
\end{eqnarray}
\begin{eqnarray}
&& - s_{ab}\; s_b\; \Bigl [ \;2 \beta \bar\beta + \bar C_\mu C^\mu - 
{\displaystyle \frac{1}{4}}
B^{\mu\nu} B_{\mu\nu} \;\Bigr ] = \bar B^\mu (\partial^\nu B_{\nu\mu})
+ B \cdot \bar B + \nonumber\\
&& \partial_\mu \bar \beta  \partial^\mu \beta
+ (\partial_\mu \bar C_\nu - \partial_\nu \bar C_\mu) (\partial^\mu C^\nu)
+ (\partial \cdot C - \lambda) \rho + (\partial \cdot \bar C + \rho) \lambda.
\end{eqnarray}
It is to be noted that the difference between the above two expressions are only
in the first term. However, modulo a total spacetime derivative, these terms are
equivalent because of the CF type restriction in (11). Thus, the Lagrangian densities
${\cal L}_B$ and ${\cal L}_{\bar B}$ are coupled (but equivalent) Lagrangian densities
for the notoph gauge theory in four dimensions of spacetime.

Due to CF type relation (11), we can have the following
expressions for the term $(B \cdot \bar B)$ that appears on the r.h.s. of 
the equations (20) and (21):
\begin{eqnarray}
B \cdot \bar B = B \cdot B - B^\mu \partial_\mu \phi, \qquad
B \cdot \bar B = \bar B \cdot \bar B + \bar B^\mu \partial_\mu \phi.
\end{eqnarray}
As a consequence of the above equations, we have the following
\begin{eqnarray}
{\cal L}_B &=& \frac{1}{12} H^{\mu\nu\kappa} H_{\mu\nu\kappa}
+ B^\mu (\partial^\nu B_{\nu\mu} - \partial_\mu \phi)
+ B \cdot  B + \partial_\mu \bar \beta  \partial^\mu \beta \nonumber\\
&& + (\partial_\mu \bar C_\nu - \partial_\nu \bar C_\mu) (\partial^\mu C^\nu)
+ (\partial \cdot C - \lambda)  \rho + (\partial \cdot \bar C + \rho) \lambda,
\nonumber\\
{\cal L}_{\bar B} &=& \frac{1}{12} H^{\mu\nu\kappa} H_{\mu\nu\kappa}
+ \bar B^\mu (\partial^\nu B_{\nu\mu} + \partial_\mu \phi)
+ \bar B \cdot \bar B + \partial_\mu \bar \beta  \partial^\mu \beta \nonumber\\
&& + (\partial_\mu \bar C_\nu - \partial_\nu \bar C_\mu) (\partial^\mu C^\nu)
+ (\partial \cdot C - \lambda) \rho + (\partial \cdot \bar C + \rho) \lambda,
\end{eqnarray}
which lead to the Euler-Lagrange equations of motion 
\begin{eqnarray}
B_\mu = - \frac{1}{2} \; (\partial^\nu B_{\nu\mu} - \partial_\mu \phi), \qquad
\bar B_\mu = - \frac{1}{2} \; (\partial^\nu B_{\nu\mu} + \partial_\mu \phi),
\end{eqnarray}
that imply the CF type condition in (11).

The coupled Lagrangian densities, that
have been derived due to the techniques of the (anti-) BRST formalism and use of the
CF type restriction (11) (emerging from the superfield formalism) are found to
be quasi-invariant under the (anti-) BRST symmetry transformations (17) and (16).
This can be seen from the following equations
\begin{eqnarray}
s_b {\cal L}_B &=& - \partial_\mu \bigl [ B^\mu \lambda + 
(\partial^\mu C^\nu - \partial^\nu C^\mu) B_\mu + \rho \partial^\mu \beta \bigr ], \nonumber\\
s_{ab} {\cal L}_{\bar B} &=& - \partial_\mu \bigl [ (\partial^\mu \bar C^\nu - \partial^\nu \bar C^\mu) \bar B_\mu 
+ \lambda \partial^\mu \bar \beta  - \rho \bar B^\mu \bigr ],
\end{eqnarray}
which establish that the action remains invariant under (16) and (17).

One would be curious to know the transformation properties of the Lagrangian density ${\cal L}_B$ under
the anti-BRST transformations $s_{ab}$ and that of ${\cal L}_{\bar B}$ under the transformations $s_b$.
It is very interesting to check that, under $s_{ab}$, the Lagrangian density ${\cal L}_B$ transforms
to a total spacetime derivative plus terms that are zero on the constrained surface defined by the field
equation (11). Similar is the situation of ${\cal L}_{\bar B}$ under the transformations $s_b$. Thus,
we conclude that the superfield formalism provides the (anti-) BRST symmetry transformations, CF type
restriction (11) and ensuing coupled Lagrangian densities for the notoph gauge theory (see, e.g. [8] and [18]).

\section{Geometrical meaning: Superfield approach}

We concisely pin-point here the geometrical meaning of the (anti-) BRST
symmetry transformations and the mathematical properties associated with them.
In fact, one can encapsulate the geometrical interpretations in the language of the
following mathematical mappings:  
\begin{eqnarray}
&&s_b \Leftrightarrow Q_b \;\Leftrightarrow \;\mbox{Lim}_{\theta \to
0} \;{\displaystyle \frac{\partial}{\partial\bar\theta}}, \qquad
s_{ab} \Leftrightarrow Q_{ab} \;\Leftrightarrow\;
\mbox{Lim}_{\bar\theta \to 0} \;{\displaystyle
\frac{\partial}{\partial\theta}}, \nonumber\\ && s_b^2 = 0
\Leftrightarrow Q_b^2 = 0 \;\;\;\Leftrightarrow \;\;\;\mbox{Lim}_{\theta \to
0}\; \Bigl ( {\displaystyle \frac{\partial}{\partial\bar\theta}}
\Bigr )^2 = 0, \nonumber\\ && s_{ab}^2 = 0 \;\;\;\Leftrightarrow \;\;\;Q_{ab}^2
= 0 \;\;\Leftrightarrow \;\;\mbox{Lim}_{\bar\theta \to 0}\;\Bigl (
{\displaystyle \frac{\partial}{\partial\theta}} \Bigr )^2 = 0, \nonumber\\
&& s_b s_{ab} + s_{ab} s_b = 0 \;\;\;\Leftrightarrow \;\;\;Q_b Q_{ab} + Q_{ab} Q_b = 0 
\;\;\;\Leftrightarrow \;\;\;\nonumber\\
&& \Bigl (\mbox{Lim}_{\bar\theta \to 0} \;{\displaystyle
\frac{\partial}{\partial\theta}} \Bigr ) \; \Bigl (
\mbox{Lim}_{\theta \to
0} \;{\displaystyle \frac{\partial}{\partial\bar\theta}} \Bigr )
+ \Bigl (\mbox{Lim}_{\theta \to
0} \;{\displaystyle \frac{\partial}{\partial\bar\theta}} \Bigr )\;
\Bigl (\mbox{Lim}_{\bar\theta \to 0} \;{\displaystyle
\frac{\partial}{\partial\theta}} \Bigr ) = 0.
\end{eqnarray}
The above (geometrically intuitive) mappings are possible only in
the super field approach to BRST formalism proposed in [13-16] where
$Q_{(a)b}$ are the nilpotent (anti-) BRST charges corresponding to $s_{(a)b}$.

The first line in (26) implies that the off-shell nilpotent
(anti-) BRST symmetry transformations $s_{(a)b}$ and their
corresponding generators $Q_{(a)b}$ geometrically correspond to
the translational generators along the Grassmannian directions of
the (4, 2)-dimensional supermanifold. To be more specific, the
BRST symmetry transformation corresponds to the translation of the
particular superfield along the $\bar\theta$-direction of the
supermanifold when there is no translation of the same superfield
along the $\theta$-direction of the supermanifold (i.e. $\theta
\to 0$). This geometrical operation on the specific superfield
generates the BRST symmetry transformation for the corresponding
4D ordinary field present in the Lagrangian densities (23). A
similar kind of argument can be provided for the existence of the
anti-BRST symmetry transformation for a specific field in the
language of the translational generator 
(i.e. $ \mbox{Lim}_{\bar\theta \to 0}\; (\partial/\partial\theta)$)
on the above (4, 2)-dimensional supermanifold.

\section{Conclusions}

It is evident that the superfield approach to BRST formalism [13,14]
is an essential theoretical tool that always leads to the derivation
of the off-shell nilpotent and absolutely anticommuting (anti-) BRST
symmetry transformations for a given 4D $p$-form gauge theory [18]. In addition, it
provides the geometrical origin and interpretation for the properties
of nilpotency and absolute anticommutativity in the language of translational
generators along the Grassmannian directions of the (4, 2)-dimensional
supermanifold. In our very recent work [21], we have been able to apply
the superfield formalism to 4D Abelian 3-form gauge theory and we have shown 
the existence of the CF type restrictions that are deeply connected
with the idea of gerbes.\\

\noindent
{\bf Acknowledgements:}
Financial support from the
Department of Science and Technology (DST), Government of India,  
under the SERC project sanction grant No: - SR/S2/HEP-23/2006, is gratefully acknowledged.  
The author is grateful to the organizers of SQS'09 for the invitation and warm hospitality at JINR.
Furthermore, he is also indebted to A. Isaev, E. Ivanov and S. Krivonos for his education
in theoretical physics at BLTP, JINR, Dubna.\\

\end{document}